# Engineering Reliable Deep Learning Systems

P. Santhanam[1], Eitan Farchi[2] and Victor Pankratius[3]

(pasanth@us.ibm.com, farchi@il.ibm.com, pankrat@mit.edu)
[1]IBM Research AI, Yorktown Heights, NY 10598
[2]IBM Research, Haifa, Israel
[3]Massachusetts Institute of Technology, Cambridge, MA 02139

**Abstract**

Recent progress in artificial intelligence (AI) using deep learning techniques has triggered its wide-scale use across a broad range of applications. These systems can already perform tasks such as natural language processing of voice and text, visual recognition, question-answering, recommendations and decision support. However, at the current level of maturity, the use of an AI component in mission-critical or safety-critical applications can have unexpected consequences. Consequently, serious concerns about reliability, repeatability, trust, and maintainability of AI applications remain. As AI becomes pervasive despite its shortcomings, more systematic ways of approaching AI software development and certification are needed. These fundamental aspects establish the need for a discipline on "AI engineering". This paper presents the current perspective of relevant AI engineering concepts and some key challenges that need to be overcome to make significant progress in this important area.

## I. Introduction

In the past decade, AI has become a part of our daily lives. Whether it is searching on Google or personal recommendations on Amazon or Netflix or Facebook, a deep learning (DL) model is working in the background. This is made even more obvious in the use of devices like Amazon/Alexa or Apple/Siri or Google/Home since they are physical manifestations of the human-machine interactions. While these systems are clearly impressive in what they have come to achieve, the consequence of mistakes by these systems are relatively minor.

Now let us consider the examples where mistakes can have a more severe social consequence: In 2016, Microsoft launched an experimental AI chat bot, called Tay. The bot 'learned' from interactions with users on Twitter and had to be shut down just a day after deployment because of its obscene and inflammatory tweets [1]. There have been many documented cases when the AI systems inherited various biases (racial, gender, age, etc.) [2]. AI is also prominent in self-driving cars. In spite of hundreds of thousands of accidents on the roads with human drivers annually, the much smaller number of accidents involving the self-driving cars get major attention [3]. It is quite likely that irrespective of the technology achievements, accepting self-driving cars on our roads is going to take major evolution in social acceptance, liability laws and government regulations [4]. Similarly, application of AI to medicine [5] also poses many technical and social challenges.

Reliability of DL systems can be severely affected by their exposure to adversarial attacks. In contrast to the cyberattacks (e.g. denial of service) commonly faced by traditional systems, the adversarial attacks can change the output behavior of the DL systems in unexpected and often subtle ways. [6]

These examples point to specific weaknesses of the DL systems due to their fundamental reliance on data and complex statistical models. As will be discussed below, the engineering of these systems differs substantially from established practices of many decades for traditional software systems. In this paper, we focus on the practical engineering aspects of building reliable software systems using DL technology available today. Many concepts also apply to broader machine learning systems. Our goal is to explain the current state of the art and identify key challenges that need to be addressed to make the integration of AI into various systems, ranging from business-critical applications to various government missions, more viable.

## II. Why AI Engineering?

### A. Traditional Software Systems

Traditional engineering practices for building complex software systems rely on the principles of functional and modular decomposition [7]. Program units must have clear boundaries and known expected behaviors at design time. A deviation from the expected behavior is the definition of a software defect (i.e. bug), which is at the heart of any quality management program. These principles are key to system lifecycle activities, i.e. design, development, testing, deployment, and maintenance. Changes in requirements, new functions, bug fixes, development activities and frequency of software releases are managed in a process (e.g., agile, waterfall, etc.) to meet business expectations. The software behavior is deterministic and the team follows good design practices such as modularity, encapsulations, separation of concerns to keep the software content tractable for humans. There are tools to support various activities (e.g. static code analysis, debugging, data flow, change management, bug tracking, test harnesses, etc.)

### B. Deep Learning Systems

The power of the AI systems lies in their ability to learn empirically and adapt incrementally to ever increasing data sets without the need for manually written programs. They perform tasks that were very difficult to realize before with high degrees of realism, because they typically did not include empirical learning in a systematic way. Several key drivers have led to this advancement, thus creating a new context for the engineering discussion. Growing data volumes in search engines, crowdsourcing, social media tagging have increased training examples for learning algorithms. Scalable computing has become an affordable commodity for everyone, thanks to multicores, GPUs, and clouds. Due to the value added by learning systems, even more specialized hardware like Vision Processing Units (VPUs) and Tensor Processing Units (TPUs) are being offered. Scalable data and hardware are thus facilitating enhancements on the learning algorithms side. As an example, combining convolution operators in neural networks allows a gradual detection and refinement of image features from the concrete towards the abstract, allowing practitioners to add refinements as needed.

These AI systems are not `engineered' in the traditional sense, as described in Section II.A. Fundamentally, there is no requirement or specification document linking inputs and outputs of functions; just training data containing examples of inputs and corresponding outputs. The AI component does the learning via complex neural networks, with no guarantees or explanations on the exact functional operations. Programs decide by data on how to behave, i.e. "data is the new specification". Testing such systems using the traditional verification techniques simply will not work, since there is no description of what the system is supposed to do. Added complexity comes from statistical nature of the machine learning algorithms which select outputs for given inputs typically based on confidence levels and hence not in a deterministic way.

Personalization of the outputs to match the specific user introduces yet another difficulty: the correctness of the output can be decided only by the subjectivity of the person looking at it, making a priori generic 'user acceptance tests' hardly practical as they can only be judged by explicit user feedback. Simply put, there is no simple definition of a bug! While the technology behind these systems is truly amazing, introducing new degrees of freedom has unintended consequences and emergent behavior that is largely invisible to developers and consumers. Additionally, the heuristic properties of programs can change over time as system behavior changes due to new patterns of relationships in the training data and continuous learning.

While the engineering of traditional software systems primarily focused on functionality, usability, reliability and operational performance, due to the potential application of AI in all walks of life, DL systems invoke a serious social concern: **trust**. The notion of trust covers a wide range of areas, such as human over-sight, robustness, data privacy, fairness, ethics, transparency and accountability [8].

Currently, DL components are built by skilled data scientists in a relatively ad hoc fashion using open source libraries with very little attention to software engineering principles. The scaling of AI to the larger ecosystem will depend critically on how these activities can be performed by more people with more engineering discipline and less skills.

### C. Triad of AI: Critical Success Factors

It is useful to think about building DL systems in terms of a triad; an interplay between three factors i.e. data, domain context and the AI algorithms (Figure 1). In this regard, algorithms play the role of a 'shell'. Operation of this shell is determined by the actual training data and the semantic labeling of the outputs, which has to make sense to the humans in the context of the specific domain or an application area. Therefore, these three aspects together define the necessary conditions and if anyone of them is missing, the resulting DL systems are likely to fail.

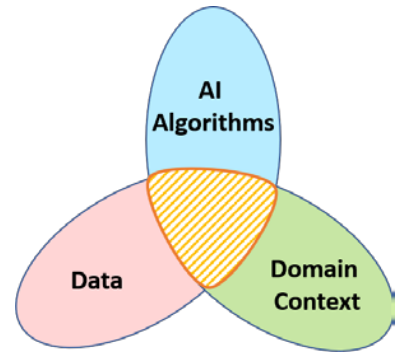

Figure 1: The "Triad of AI": Critical Success Factors

As an example, let us consider a mortgage approval system. The system has to rely on learning algorithms for risk assessment, customer data with personal information, and domain knowledge on laws, financial guidelines and regulations. This interplay might be different in various countries as regulations differ. For example, European Union's General Data Protection Regulation (GDPR) [9] requires that customers have a right to know how their personal data is used, which can include decisions made by AI. The legal framework also provides constraints to what data can be collected in the first place, which has an influence on the algorithmic workings of learning systems.

## III. Software Systems for the Government

Governments have unique responsibilities in serving the public and their software systems have to be sensitive to wide range of concerns. National Aeronautics & Space Agency (NASA) defines [10] "mission critical" as loss of capability leading to possible reduction in mission effectiveness" and "safety-critical" means failure or design error could cause a risk to human life. These definitions can be broadly interpreted to provide guidelines for both civilian and national defense missions. In the civilian space, there are many departments that can benefit from AI such as Treasury, Justice, Health & Human Services, Labor, etc. These departments have strong requirements for record keeping, fairness, data privacy, transparency of decisions and audit of adherence to standard processes. As for data usage, GDPR [9] puts severe restrictions on the use of individual data for building AI models and their maintenance.

In 2018, US Department of Defense published [11] their AI strategy with five strategic focus areas. One of them was "Scaling AI's impact across DoD through a common foundation that enables decentralized development and experimentation" that includes "shared data, reusable tools, frameworks, and standards, and cloud and edge services". In another report from the Office of the Director of National Intelligence (ODNI) [12], on Augmenting Intelligence using Machines (AIM) initiative, explicit discussion of AI assurance included the importance of data engineering, the need for a robust and sustainable software, initial and continuous performance evaluation to match the mission goals, the need for rigorous testing regimes, concept drift in models and dealing with possibilities of adversarial attacks on models and data.

A key concept in the deployment of software systems in the US government agencies is the use of Technology Readiness Levels (TRL) [13] to assess the maturity of the systems for mission use. Software components also require certification under Federal Information Security Management Act of 2002 (FISMA) as defined by the National Institute of Standards and Technology Special Publication 800-53 [14]. To complicate matters further, there are various clearance levels for security and data, and the related AI artifacts have to be dealt with appropriately. There may also be clearance levels required for people and facilities to execute the project that can place constraints on skill levels of staff and their roles.

In either civilian or national defense context, the AI has to fit into a larger system with multiple endpoints and established processes. The need for low barrier for Human-Machine collaboration is paramount. An unexpected or inappropriate behavior of AI supporting a government mission can be disastrous.

## IV. AI Engineering Lifecycle Activities

For the foreseeable future, mission critical systems may need one or more DL components along with traditional AI functions (e.g. planning, rule based), non-AI functions (e.g. traditional analytics, reporting) and IT components (e.g. access control). Figure 2 shows the various engineering lifecycle activities for an application with one DL component that supports a task required for a mission (e.g. object identification). There are two overlapping sets of activities: one set for the application lifecycle supporting the mission (circle on the left) and another set for the DL component supporting the application (circle on the right), with the circle in the middle representing the intersection between the two. The DL model lifecycle may go through many more iterations before delivering a model suitable to support the mission task. Project planning has to include these aspects when deciding on resources and schedules. This section describes these activities and highlights the impact of DL components.

### A. Application Lifecycle

The AI application lifecycle consists of traditional software construction activities, *except that it is influenced by the inclusion of the DL component(s) for one or more tasks*. Going clockwise, the activities and their purpose (starting top left in Figure 2) are as follows:

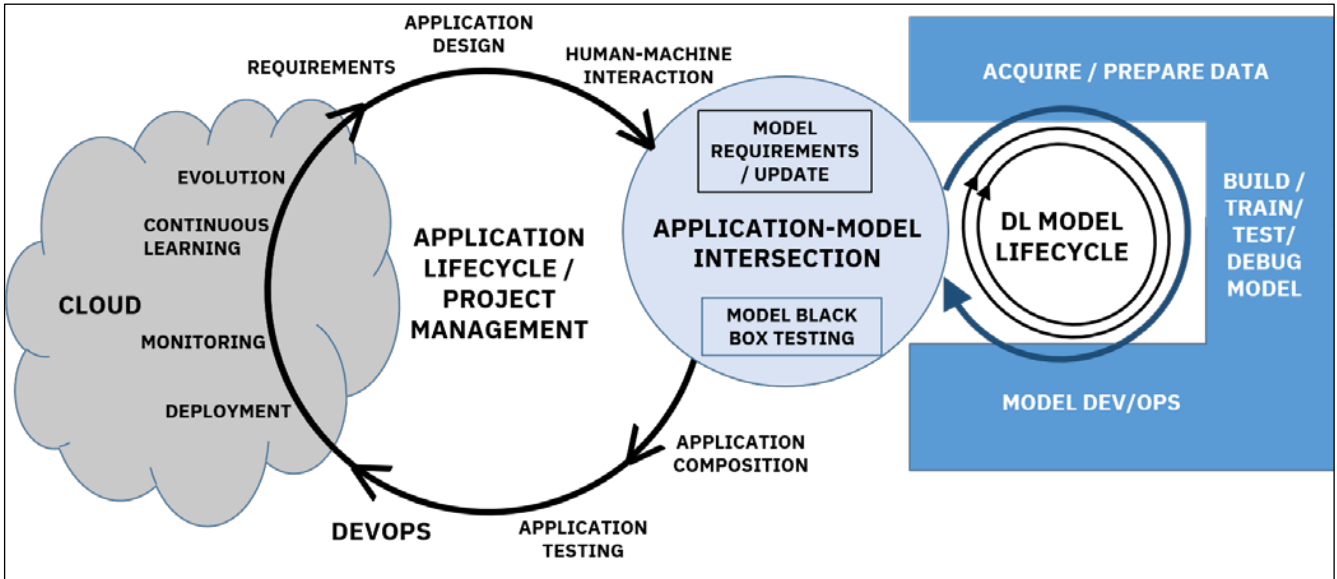

Figure 2: Engineering lifecycle with one deep learning component

**Application Requirements:** These requirements are generally determined by the mission goals and system expectations and are independent of choices on AI. It may be worthwhile to capture specific concerns about robustness, security, bias, ethics, human level explanation and system transparency explicitly. Such constraints will help avoid the inappropriate use of DL components.

**Application Design:** This activity requires careful consideration of which task in the application can be reliably executed by a DL component. If the output of the DL task has a high consequence (e.g. human life) and the confidence in the DL output is low, then the task is not suitable for AI. Conversely, if the output has low consequence and high confidence, then it is an ideal task for AI. A critical requirement is the availability of data of adequate quality and quantity [15] before a DL model building process is attempted. Rule-based checkers can help monitor DL component outputs and their impact on mission or safety objectives.

**Human-Machine Interaction:** System usage model moves from "humans using the machines" to one of "human-machine collaboration" (i.e. augmenting human intelligence), thus impacting human experience and productivity directly. AI based applications also leverage additional forms of human input (i.e., unstructured text, gestures, speech). The decision on which interaction paradigms to use depends on the quality of the machine cognition of the human input [16] as well as on the ability for a graceful recovery in the event of its cognitive failures.

**Application-Model Intersection:** There are two roles for this activity: (1) Defining model requirements and expected quality metrics such as accuracy, runtime performance, robustness, explainability, etc. and (2) Create black box testing to verify if the model indeed satisfies the model requirements. These activities need to be carried out by persons with a good understanding of the mission requirements. To test the model outputs, separate data sets (i.e. hold-out sets) are required that are not actually used in the model building process [17] and that the data distribution in the hold-out set represents the mission requirements. If the model does not meet the mission requirements, the development process has to go back to the DL model lifecycle activities for improvements.

**DL Model Lifecycle:** Current best practices in creating machine learning applications are discussed in [17, 18]. Important activities are:

- *Acquire / prepare data*: Raw data acquisition may involve licensing, security, privacy issues as well as proper governance after the acquisition. DL modeling currently requires large amounts of labeled data that may have to be acquired from domain experts or crowd sourcing at considerable cost. As is known in the statistics community, nearly 80% of the effort in a data project is related to data preparation. For AI applications, proper preparation is needed to avoid bias and ensure fairness and trust. Feature extraction is a critical task in this process, which can help remove redundant data dimensions, unwanted noise, and other properties that degrade model performance. This topic will be revisited in Section V.

- *Build/Train/Test/Debug Model*: This step aims to produce the best model meeting the mission requirements with the available data. In practice, various programming frameworks (e.g. TensorFlow, PyTorch, Scikit-Learn, etc.) are used to create the model code. These frame-works typically provide some tool support for the coding process. However, as should be evident, *even if the model code does not have any errors in it, that does not mean that the model is good for the mission purpose.* Another important step in building DL models is the separation of training data and validation data so that the model's ability to generalize can be evaluated accurately [17]. Cross-validation is a standard practice, where typically 70% of the data is used for training and 30% of the data is used for validation. This helps to tune model parameters, select data features, and tweak the learning algorithm. The data needs to be drawn from the same distribution for the training and validation sets. Unfortunately, debugging of the DL models is complex [19] since DL behavior can be the result of the model-inferred code and the underlying training data. Breck et al [20] discussed 28 specific tests and monitoring needs, based on experience with a wide range of production ML systems at Google.
- *Model Dev/Ops*: When the model is deemed adequate, it has to be integrated into the actual software application. Black box testing evaluates if the model is "good enough" for deployment or suggests a lifecycle restart. The Dev/Ops activity also needs to keep the model and training data versions in sync so that any changes to the model in the future can be adequately tracked.

**Application Composition:** Tested DL components are integrated into applications as black box services, much like other micro-services in use today. This task may combine other non-DL services to create a complete software application product.

**Application Testing:** Application level testing is similar to the User Acceptance Testing in traditional software development. However, it has a few twists: There are specific challenges involving testing for any system behavior resulting from customizations and automated sensing of an individual's profile. Evaluators need to execute the application in relevant target environments with their resource constraints (e.g. portable devices on the edge, etc.). Results on robustness, security, bias, ethics, system transparency etc. need to be documented and addressed.

**Application Deployment/DevOps:** The absence of the concept of a defect in AI systems challenges traditional notions of success and failure in integration testing and deployment. In addition, data properties fed into DL systems can change after deployment. Thus, new quality assessment techniques are required for such contexts. We discuss this further in Section V.

**Monitoring:** In traditional software systems, monitoring had mainly two purposes: (i) understanding user behavior to improve system design (e.g. A/B testing) (ii) anticipating potential performance issues that require maintenance upgrades. However, many systems can function quite adequately without such monitoring. By contrast, *in AI systems monitoring is not an option, but a required activity.* This is because DL model behavior can drift with time due to changes in the input data and the data distributions in the deployment. The specific details on what to monitor and how often will depend on the specific use cases and mission goals. We return to discuss drift further in Section V.

**Continuous Learning & Evolution:** This activity aims to modify the DL model to match the mission requirements as data assumptions change. One common challenge with this goal is that the data collected during deployment is not 'labeled' (i.e. the correct output is unknown) and hence not directly usable for model re-training. Techniques such as active learning [21] can be used to get users to provide the labels directly; semi-automated approaches combine automated labeling and human validation. As Microsoft Tay bot example [1] demonstrated, continuous model learning has its own challenges in validation.

## V. Current Challenges in AI Engineering

Various authors have addressed the topic of engineering challenges in DL systems. A recent summary is presented by Khomh et al. [22]. Masuda et al. [23] performed a literature survey on key problems faced by machine learning applications and identified potential software engineering approaches to solve these problems.

### A. Related Work

**Technical debt.** Skully et al. [24] concluded that DL applications carry significant technical debt. Due to lack of clear abstraction boundaries with specific intended behaviors, the behavior of a machine learning component can be summarized as "Change Anything - Changes Everything". Data dependencies are more costly than code dependencies. Systems can also become overly complex due to glue code needed to support various models, data processing, and ubiquitous experimentation. Maintenance is expensive due to inevitable changes in data and models with time.

**AI Development is different.** Amershi et al. [25] reported on a study of software teams at Microsoft as they develop AI-based applications. They concluded with three key observations: (1) managing the AI data life cycle is harder than other types of software engineering, (2) model customization and reuse require very different skills that are not typically found in software development teams, and (3) AI components are more difficult to handle than traditional software components due to the difficulty of isolating error behavior. Another study at Microsoft by Kim et al. [26] looked, in detail, at the technical and cultural challenges of data scientists being a part of software development teams.

**Tracking Complexity.** Arpteg et al. [27] studied seven AI projects and corroborated many of the observations already made in this paper and related work. They high-lighted the difficulties in tracking the various experiments with contextual information and the dependencies across the various hardware and software components. They also point to the practical problem of estimating the effort needed to build an acceptable DL model for project planning.

**Maturity and scalability.** Akkiraju et al. [28] presented an AI maturity model and a set of best practices from client experiences of building large scale real-world machine learning models at IBM.

## B. Managing DL Performance Drift

Let us consider the example of identifying objects (e.g. cats) in images with a DL component. Based on the results of our testing on the training set, (say) we expect an accuracy of identifying a cat in the interval [89%, 90%] with a probability of 99%. However, at deployment time we find out that the actual accuracy "in the wild" is 70%, indicating that a drift has occurred.

There is a difference between model drift and data drift [29]. Data drift denotes a change in the data distribution during deployment compared to the distribution during the training phase of the DL model. The basic theory of machine learning (e.g. Probably Approximately Correct learning [30] ) does not expect a data drift. While the data drift is necessary for the model performance drift, but it is not sufficient. It is quite possible that a DL model is not sensitive to the changes in the data distribution. Such a behavior is indeed, desirable. In fact, one can design the DL model to have more stable behavior by introducing anticipated data changes into the design of the neural network [31]. In our example, if we had enough instances of white and black cat images in the training set, the DL model accuracy may become immune to changes in the proportion of white and black cats at the deployment time.

How quickly and accurately can the model output drift be detected when data drift occurs? The answer to this question is complicated by the following three factors: (i) *The curse of high dimensions.* Since typical data used in learning has high dimensions, finding the difference between two distributions is a hard problem. (ii) *Data at the deployment time is typically not labeled.* As a result, direct measurement of the DL model performance is not possible. (iii) *Repeated measurement of the DL performance requires care*. Typically, we need to use proper experimental design that requires advanced sequential test analysis or stopping time techniques to retain statistical power.

In short, to avoid model drift issues, one has to rely on proper engineering. In the model building process, clear separation of training and validation data which are sampled from the same distribution, is necessary. In addition, careful selection of black box test data is critical, so that it represents the expected data at deployment time accurately.

## C. Managing the Data Workflow

Traditional software development has mature processes to capture relevant artifacts (i.e. requirements, designs, code versions, test cases, deployment data, etc.) for future use. In contrast, the process for developing DL components may not allow the permanent storage of the training data as a critical artifact. In practice, this may be due to very large data sets, limited data access, or contract/license terms that prohibit data use beyond the training period. Data can also vary wildly over time (e.g. Twitter data) making the data capture less relevant for later use. If it is impractical to capture and preserve all the data used to train and test an AI system, developers and auditors will be unable to reproduce or post-audit the models and components that are vital to the operation of mission critical AI applications.

Data preparation is the most underrated task in DL model building. It holds the key to a better and more reliable model building. Data preparation affects the final predictive accuracy, generally even more than the actual modeling step! Like modeling, it also contains parameters which should be tuned. Currently, data preparation for DL is a "black art" giving rise to many conceptual errors in practice. Data preparation must be understood as a process that must be optimized, cross-validated, and deployed jointly with data modeling in order to ensure proper applicability.

Figure 3 represents a typical data workflow [32] in a deep learning project. The provenance of the data sources and their trustworthiness are not typically questioned by the data scientists. They often make ad hoc assumptions and data transformations in the process, which are not recorded for reproducibility. Thus, tools are critical to help with the data workflow tasks shown in Figure 3. Gradual progress is being made in this area. For example, Snorkel [33], is an opensource system to build and manage training datasets programmatically. It currently focuses on three key operations: labeling data (e.g. using heuristic rules or distant supervision techniques), transforming data (e.g. data augmentation and capturing invariances) and slicing data into different critical subsets. Khurana et al. [34] used reinforcement learning on past examples to learn a policy across data sets to automate feature engineering efficiently. With One Button Machine, Lam et al. [35] automated fea-ture discovery in relational databases with impressive performances in Kaggle competitions. More examples to automate data workflow can be found in Ref. [32].

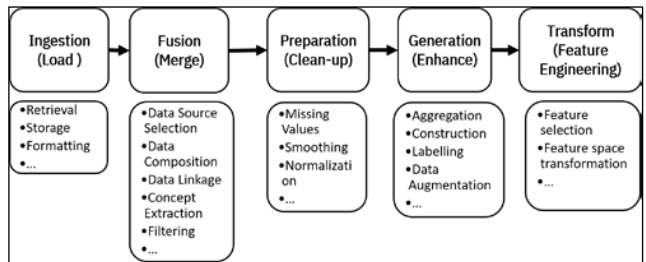

Figure 3: Typical Data Workflow

### D. Ensuring Model Correctness Across the Lifecycle

In the absence of a clear definition of a software defect, how do we manage the correctness of a DL system across the software lifecycle? Hestness et al. performed a careful empirical study [36] for a few different datasets and found that the error in generalization of DL models decreased as a power law with the training data size. Kendal and Gal [37] used an alternative highly theoretical Bayesian approach to calculate uncertainties in DL models, *since the deep learning models themselves are not able to represent uncertainty* and they do not deal with the lifecycle aspects. Perera et al. [38] addressed the need for analyzing and preparing the data to match the accuracy of the expected model accuracy.

One potential path is to augment the deterministic approach for measuring the correctness in traditional software with a statistical approach. Since DL is a statistical technique, the measurements have to reflect that property. Thus, techniques similar to Statistical Process Control (SPC), commonly used in manufacturing, should be considered for DL systems. As with traditional software development, it is important to define a consistent set of metrics across all the activities of a DL application to track the correctness of the application as it moves through the lifecycle. These metrics have to capture the variance of the outputs for given inputs, as well as the variance of the outputs for different input data sets in a normalized manner to facilitate meaningful comparisons. The overall variance of the DL model outputs is likely to be a function of the variance of the input data distributions and the specific transformations performed by the DL models. We need a way to capture this variance in a practically useful and consistent manner across the lifecycle activities.

### E. AI Certification and Benchmarking

Component-based software engineering has become the norm in building software systems, achieving significant reuse. Given the abundance of AI services and components (i.e. algorithms, libraries, frameworks) available from open source and commercial offerings, how can the government (for that matter, any one) absorb the technology easily and quickly to support mission-critical needs? Particularly, open source components come with 'use as is' terms, clearly implying significant risk to the user. Then, there is the matter of 'Trust' in AI systems we mentioned earlier. The purpose of this topic here is to provoke a discussion on what dimensions capture the risk of the user and how to mitigate that by some manner of certification acceptable to the community at large. The concept of a Fact Sheet [39] is very appealing and much like food labeling, can provide information on the various dimensions that matter to the user. For example, these can include accuracy (e.g., F1 score for classification tasks), fairness (bias quantification in the model outputs for certain input data distributions), lineage (representation of provenance of the data used to train the model and clarity over the ownership of the model), or robustness (a measure of resistance to adversarial attacks).

As deep learning is data driven, AI component certification will also require standard data benchmarks and competitions in the community in order to be able to compare component metrics on common baselines. Setting up community/industry-wide data benchmarks for AI will be a necessity that compares to past efforts in benchmarking CPUs (e.g., SPEC), parallel computers (e.g., TOP500, PARSEC, parallel I/O), SAT solving, chess, and other aspects. The results of AI benchmarks will facilitate an understanding of the Pareto frontiers that show how components perform in various dimensions, what their "Technology Readiness Level" is, what their individual tradeoffs are, and how well their internal functionality complies with regulations (e.g. GDPR, FISMA).

## VI. Conclusion

This paper explores the topic of AI engineering required to build reliable deep learning-based software systems. We discuss the impact of the introduction of deep learning components on the traditional software lifecycle activities. Considerable challenges remain in the scaling of already proven deep learning technology to real-world systems that are mission-critical or safety-critical. We identify four key engineering areas that need breakthroughs: (i) techniques for managing deep learning model performance drift during deployment (ii) tool support to improve the data workflow tasks in the model building process (iii) methods to measure the correctness of models across their lifecycle that augments or replaces the current 'defect' based quality management system for software (iv) certification processes that will enhance the adoption of commercial or open source components into mission critical applications. With additional progress in these areas, AI can become a more trustworthy partner in mission critical applications.

## Acknowledgments

Authors thank R. Akkiraju, K. Bhaskaran, G. Booch, R. Chang, B. Hailpern, M. Hind, O. Raz, C-F. Shu, J. Smith and F. Stein for useful discussions and inputs. Victor Pankratius acknowledges support from NSF ACI1442997 and NASA AISTNNX15AG84G.